\documentclass[journal]{IEEEtran}
\ifCLASSOPTIONcompsoc
\else
\fi
\ifCLASSINFOpdf
\else
\fi
\usepackage{graphicx}
\usepackage{amssymb}
\usepackage{amsmath}
\usepackage{bm}
\usepackage[Algorithm, boxed]{algorithm}
\usepackage{algorithmic}
\usepackage[bookmarks=false]{hyperref}
\usepackage{breakurl}
\usepackage{multirow}
\begin{document}
%
\title{Analytical Bounds between Entropy and Error Probability in Binary Classifications}
%
%
%
%

\author{Bao-Gang Hu,~\IEEEmembership{Senior Member,~IEEE},~Hong-Jie Xing
\IEEEcompsocitemizethanks{\IEEEcompsocthanksitem Bao-Gang Hu is with NLPR/LIAMA,
Institute of Automation, Chinese Academy of Sciences, Beijing 100190, China.\protect\\
E-mail: hubg@nlpr.ia.ac.cn \protect\\
   Hong-Jie Xing is with College of Mathematics and Computer Science, HeBei University,
Baoding, 071002, China. \protect\\
E-mail: hjxing@hbu.edu.cn 
}
\thanks{}}

\IEEEcompsoctitleabstractindextext{%
\begin{abstract}
The existing upper and lower bounds between entropy and error probability 
are mostly derived from the inequality of the entropy relations, 
which could introduce approximations into the analysis. 
We derive analytical bounds
based on the closed-form solutions of conditional entropy 
without involving any approximation. 
Two basic types of classification errors are investigated
in the context of binary classification problems, namely, 
Bayesian and non-Bayesian errors.
We theoretically confirm that Fano's lower bound is an exact lower bound for any types 
of classifier in a relation diagram of ``error probability vs. conditional entropy''.
The analytical upper bounds are achieved with respect to the minimum prior probability, 
which are tighter than Kovalevskij's upper bound.

\end{abstract}

}

\maketitle

\IEEEdisplaynotcompsoctitleabstractindextext

%
\IEEEpeerreviewmaketitle

\section{Introduction}
\label{sec:introduction}
In information theory, the relations between entropy and error 
probability are one of the important fundamentals.  
Among the related studies, one milestone is Fano's inequality (also known as Fano's 
lower bound on the error probability of decoders),
which was originally proposed in 1952 by Fano,
but formally published in 1961 \cite{Fano1961}.
It is well known that Fano's inequality plays a critical role in 
deriving other theorems and criteria in information theory 
\cite{cover2006}\cite{verdu1998}\cite{yeung2002}. 
However, within the research community,  it has not been widely accepted exactly 
who was
first to develop the upper bound on the error probability \cite{Golic1999}. 
According to \cite{Vajda2007} \cite{Morales2010},
Kovalevskij \cite{Kovalevskij1965} was possibly the first 
to derive the upper bound of the error probability in relation to entropy
in 1965. Later, several
researchers, such as
Chu and Chueh in 1966 \cite{Chu1966}, Tebbe and Dwyer III 
in 1968 \cite{Tebbe1968},
Hellman and Raviv in 1970 \cite{Hellman1970},
independently developed upper bounds.    

The upper and lower bounds of error probability 
have been a long-standing topic in studies on
information theory \cite{Chen1971} \cite{Bassat1978} \cite{Golic1987} \cite{Feder1994}
\cite{Han1994} \cite{Poor1995} \cite{Harremoes2001} 
\cite{Erdogmus2004}\cite{Vajda2007} \cite{Morales2010}. 
However, we consider two issues that have received less 
attention in these studies:

I. What are the ``{\it analytical bounds}'' for
which approximations have not been applied in the derivation?

II. What is the interpretation of each bound or some key points in a given 
diagram of entropy and error probability? 

On the first issue, we define ``{\it analytical bounds}'' to be those derived
from closed-form solutions, rather than from inequality approximations.
Generally, exact bounds are desirable from the viewpoint of theory 
and applications.
The second issue suggests the need for a better understanding of
the bounds in the relation of entropy
and error probability. For example, 
some key points located at the
bounds could show the specific interpretations for theoretical insights or 
application meanings.  

The above issues forms the motivation behind this work.
We establish analytical bounds based on
closed-form solutions. 
Furthermore, we study the 
bounds in a wider range of error type, i.e., Bayesian and 
non-Bayesian. Non-Bayesian errors are also of importance 
because most classifications are realized within this category. 
We take classifications as a problem background
since it is more common and understandable from our daily-life experiences. 
We intend to simplify settings within binary states and Shannon entropy
definitions so that the analytical-principle approach is highlighted.
Based on this principle, one is able to extend the study to more general 
classification settings, such as multiple-class (or multihypothesis) problems, 
and on other definitions of entropy, 
such as R\'{e}nyi entropy.

The rest of this paper is organized as follows. In Section II, we
present related works on the bounds. For a problem background
of classifications, several related definitions are given in Section III. 
The analytical bounds are given and discussed for Bayesian
and non-Bayesian errors in Sections IV and V, respectively.
Interpretations to some key points are presented in 
Section VI. Finally, in Section VII we conclude the work and present
some discussions.

\section{Related Works}
Two important bounds are introduced first, which form the baselines for the
comparisons with the analytical bounds. 
They were both derived from inequality conditions\cite{Fano1961}\cite{Kovalevskij1965}. 
Suppose the random variables $X$ and $Y$ representing input and output messages 
(out of $m$ possible messages), 
and the conditional entropy $H(X|Y)$ representing the average amount of information lost
on $X$ when given $Y$.
Fano's lower bound \cite{Fano1961} is given in a form of:
\setcounter {equation} {1}
$$ H(X|Y) \leq  H(e) +P_e {log_2 (m-1)},  \eqno (1) $$
where $P_e$ is the error probability, and $H(e)$ is the associated binary 
Shannon entropy defined by \cite{shannon1948}:
$$ H(e) = - P_e log_2 P_e - (1- P_e)log_2 (1-P_e). \eqno (2) $$
The base of the logarithm is 2 so that the units are ``{\it bits}''.

The upper bound is given by Kovalevskij \cite{Kovalevskij1965}
in a piecewise linear form:
\setcounter {equation} {2}
\begin{equation}
\label{equ:}
\begin{array}{r@{\quad}l}
&  H(X|Y) \geq  log_2 k +k(k+1)(log_2 \frac{k+1}{k})(P_e - \frac{k-1}{k}), \\
& ~~~~~~~~~~~~~~~~~~~~~~~~~~~~~~~~~~~~ and ~ ~k < m,~ m \geq 2. 
\end{array} 
\end{equation}

For a binary classification ($m=2$), Fano-Kovalevskij bounds become:
$$ H^{-1}(e) \leq P_e \leq \frac{H(X|Y)}{2}, \eqno (4) $$
where $H^{-1}(e)$ is an inverse of $H(e)$.

Several different bound diagrams between error probability
and entropy have been reported in literature. The initial
difference is made from the entropy definitions, such as Shannon
entropy in \cite{Chen1971}\cite{Golic1987}\cite{wang2008}\cite{hu2012},
and R\'{e}nyi entropy in \cite{Feder1994}\cite{Vajda2007}\cite{Morales2010}.  
The second difference is the selection of bound relations, such
as ``$P_e$ vs. $H(X|Y)$'' in \cite{Chen1971}, ``$H(X|Y)$ vs. $P_e $'' 
in \cite{Golic1987} \cite{Feder1994}\cite{Vajda2007} \cite{Morales2010},
``$P_e$ vs. $MI(X,Y)$'' in \cite{Eriksson2005}, and
``$NMI(X,Y)$ vs. $A$'' in \cite{wang2008}, 
where $A$ is the accuracy rate, $MI(X,Y)$ and $NMI(X,Y)$ are the mutual information and
normalized mutual information, respectively, between variables $X$ and $Y$.
Wang and Hu \cite{wang2008}
was the first to derive the analytical relations of mutual information
with respect to accuracy, precision, and recall, and their analytical bounds. 
However, they did not consider the Bayesian error constraint.
When the Bayesian error constraint was added into the bound
relation in \cite{hu2012}, the upper bound from \cite{Kovalevskij1965} 
is not analytical one. Because the existing bounds are derived from 
inequality with approximations, 
some investigations \cite{Taneja1985} \cite{Poor1995}
\cite{Erdogmus2004} have been reported on
the improvement of bound tightness.

\section{Related Definitions}

Binary classifications are considered in this work. 
A theoretical derivation of
relations between entropy and error probability, 
is achieved based on the joint probability $p(t,y)$
in classifications, where $t \in {T} = \{t_{1}, t_{2}\}$ 
is the true (or target) state
within two classes, and $y \in {Y} = \{y_{1}, y_{2}\}$ is the classification output.
The simplified notations for $p_{ij}=p(t,y)=p(t=t_i,y=y_j)$ are
used in this work. 
Several definitions are given below.

{\bf Definition 1} {\it (Joint probability in binary classifications):} 
In a context of binary classifications, the joint probability $p(t,y)$
is defined in a generic setting as:
$$\setlength\arraycolsep{0.1em}
 \begin{array}{r@{\quad}l}
  & p_{11}=p_1-e_1, ~  p_{12}=e_1,\\
   & p_{21}=e_2, ~~~~~~~  p_{22}=p_2-e_2, 
 \end{array}
\eqno (5) 
$$
where $p_1$ and $p_2$ are the prior probabilities of Class 1 and Class 2, respectively;
their associated error probabilities are denoted by $e_1$ and $e_2$, respectively.
For the Bayesian decision, $p_1$ and $p_2$ are always known.  
The constraints in eq. (5) are given:
$$\setlength\arraycolsep{0.1em}
 \begin{array}{r@{\quad}l}
& 0 < p_1 <1, ~~ 0 < p_2 <1, ~ p_1+p_2=1 \\
& 0 \leq e_1 \leq p_1, ~ 0 \leq e_2 \leq p_2. 
\end{array} 
\eqno (6) 
$$

{\bf Definition 2} {\it (Bayesian error and non-Bayesian error):} 
``{\it Bayesian error}'' is defined to be the theoretically lowest 
error in classifications \cite{duda2001}, and denoted by $P_e$.
Hence, the other errors are ``{\it non-Bayesian errors}'', and denoted by $P_E(>P_e$ 
for the same probability distributions). 

{\bf Definition 3} {\it (Error probability calculation):} 
In binary classifications, error probabilities 
are calculated from the same formula: 
$$ e (P_e, ~ or~ P_E)= p_{12}+p_{21}, \eqno (7) $$
where $e$ is also denoted an error variable 
with no distinction between error types. 

{\bf Definition 4} {\it (Minimum and maximum error bounds in 
classifications):} 
Classifications suggest the minimum error bound as:
$$ (P_E)_{min} = (P_e)_{min} = 0, \eqno (8) $$
where the subscript ``{\it min}'' denotes the minimum value.
The maximum error bound for Bayesian error in binary classifications is 
\cite{hu2012}:
$$ (P_e)_{max} = p_{min} =min \{p_1,p_2 \}, \eqno (9) $$
where the symbol ``{\it min}'' denotes 
a ``{\it minimum}'' operation.
For non-Bayesian error, its maximum error bound
becomes 
$$ (P_E)_{max} = 1. \eqno (10) $$

{\bf Definition 5} {\it (Admissible area, point and their properties
in a diagram of entropy and error probability):} 
In a given diagram of entropy and error probability, 
we define the area enclosed by the bounds
to be ``{\it admissible area}'', if every point
inside the area can be possibly realized from classifications. 
we call those points to be ``{\it admissible points}''.
If a point is unable to be realized from classifications, 
it is a ``{\it non-admissible point}''.
A non-admissible point can only be located
at or outside the boundary of the admissible area.
If every point located on the boundary is admissible, we call
this admissible area ``{\it closed}''. If one or more points  
at the boundary are non admissible, the area is said ``{\it open}''.

\section{Analytical upper and lower bounds for Bayesian errors}

All analytical bounds are derived from a closed-form relation
of conditional entropy and error probabilities (see Appendix A).
The analytical lower bound for Bayesian errors is:
\setcounter {equation} {10}
\begin{equation}
\label{equ:11}
P_e \geq max \{ 0, G_1(H(T|Y)) \},
\end{equation}
where $H(T|Y)$ is the conditional entropy for the random variables $T$ and $Y$,
and $G_1$ is called the ``{\it analytical lower bound function}'' 
(or ``{\it analytical lower bound}'' for short) 
and satisfies the following relations with respect to the error variable $e$: 
\begin{equation}
\label{equ:12}
\begin{array}{r@{\quad}l}
& e = G_1(H(T|Y))=H^{-1}(e), ~ and  \\
& H(T|Y)=G_1^{-1}(e)=H(e)  \\
& ~~~~~~~~~~= - P_e log_2 P_e - (1- P_e)log_2 (1-P_e). 
\end{array} 
\end{equation}
The analytical upper bound is given by:
\begin{equation}
\label{equ:3}
P_e \leq min \{ p_{min}, G_2(H(T|Y)) \}, 
\end{equation}
where $G_2$ is the ``{\it analytical upper bound function}'' 
and for which the following relation holds: 
\begin{equation}
\label{equ:4}
\begin{array}{r@{\quad}l}
& H(T|Y)=G_2^{-1}(e)  \\
& =-p_{min}log_2 \frac{p_{min}}{e+p_{min}}-elog_2 \frac{e}{e+p_{min}}.
\end{array} 
\end{equation}
In eq. (14) $p_{min}$ is known, because 
for Bayesian classifications $p_1$ and $p_2$ are given information.

Fig. 1 depicts three analytical upper bounds together with 
Fano's lower bound and Kovalevskij's upper bound 
in the graph of ``$P_e$ vs. $H(T|Y)$''.
Several findings can be observed from the novel upper bounds. 
  
\begin{figure}
\centering
\includegraphics[width=3.3in]{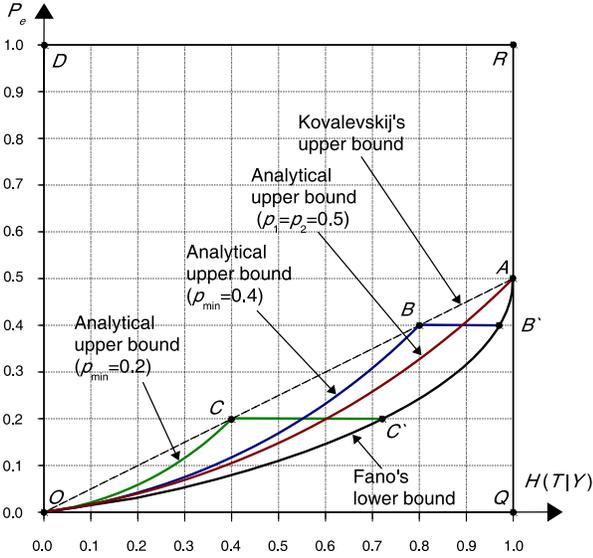}
\caption{Plot of ``$P_e$ vs. $H(T|Y)$'' giving 
the analytical upper bounds, Kovalevskij's 
upper bound and Fano's lower bound.
}
\label{fig:1}
\end{figure}
 
I. If $p_1 \neq p_2$, the analytical upper bounds are formed by 
one curve and one line. These are lower than Kovalevskij's upper bound
except for two specific points: the original point, $O$, and 
one corner point, $B$ or $C$, in Fig. 1.

II. If $p_1 = p_2$, the analytical upper bound becomes a single curve, which
is also lower than Kovalevskij's upper bound, except at the two end-points,
points $O$ and $A$.   

III. The analytical upper bounds,
either curved or linear, are controlled by $p_{min}$.   

IV. The admissible area in Bayesian decision is closed. Its shape
changes depending on the value of $p_{min}$.
For example, the area enclosed by the two-curve-one-line 
boundary, ``$O-C-C'-O$'' in
Fig. 1, corresponds to classifications with $p_{min}=0.2$.
The line boundary shows the maximum error for Bayesian
decisions, $(P_e)_{max}=p_{min}$, in binary classifications \cite{hu2012}.

Interpretations are given below to the
analytical bounds in the context of binary classifications. 
Similar discussions on some specific points
are gvien in Section VI. 

{\it Fano's lower bound}: 
In \cite{cover2006}, a marginal probability distribution is applies
for explaining the equality of Fano's lower bound (see eq. (2-144), \cite{cover2006}):
\begin{equation}
\label{equ:5}
\begin{array}{r@{\quad}l}
p(y)= (1-P_e, \frac {P_e} {m-1}, ..., \frac {P_e} {m-1}).
\end{array} 
\end{equation}
Because we derive the bound based on joint probability distributions in (5),
novel explanations can be obtained.
A generic classification setting can represent this bound:
\begin{equation}
\label{equ:5}
\begin{array}{r@{\quad}l}
e_1= \frac {p_1(p_2-e_2)} {p_2}, ~or~ e_2= \frac {p_2(p_1-e_1)} {p_1},
\end{array} 
\end{equation}
The setting above is derived based on the minimum 
relations (or Property 7 in \cite{hu2008}). Eq. (16) describes an extremal property
in the relations of entropy and error probability, 
but is expressed between the error probabilities. 

Based on eq. (16), a specific classification setting
can be obtained, in which one is 
to classify a minority class (say, Class 2) 
into a majority class (Class 1): 
\begin{equation}
\label{equ:6}
\begin{array}{r@{\quad}l}
& p_{11}=p_1, ~~~~~~ p_{12}=0, ~~~~~~~~\\
& p_{21}=p_2=e, ~  p_{22}=0.
\end{array} 
\end{equation}
Eq. (17) will result in a zero value for the mutual information, 
which implies ``{\it no correlation}''  \cite{duda2001}
between two variables $T$ and $Y$, or ``{\it zero information}''  \cite{mackay2003}
from the classification decisions. 
It also indicates the ``{\it statistically independent}'' \cite{cover2006} between 
two variables. In \cite{hu2012}, Hu demonstrated that Bayesian classifiers
will obtain such solutions for $p_1>p_2$ when processing extremely-skewed
classes with no cost terms given.  One can also observe that eq. (17)
is equivalent to (15) when $m=2$.

{\it Analytical upper bound}: 
Supposing $p_1 > p_2$, a specific classification setting 
can be obtained for representing this bound:
\begin{equation}
\label{equ:7}
\begin{array}{r@{\quad}l}
& p_{11}=p_1-e_1, ~  p_{12}=e_1=e,  \\
& p_{21}=0, ~~~~~~~~  p_{22}=p_2. 
\end{array} 
\end{equation}
Eq. (18) suggests 
the generic conditions, $e_i=e,$ if $p_i > p_j,$ and $i\neq j, ~ i,j=1,2$,
for another extremal property in the relations of
entropy and error probability. Hence, the analytical upper bound function
corresponds to a zero value for the conditional probability,
or the maximum value for the mutual information. 

\section{Analytical upper and lower bounds for non-Bayesian errors}

In a context of classification problems, Bayesian errors
can be realized only if one has exact
information about all probability distributions \cite{duda2001}. 
The assumption above is generally impossible
in real applications. Therefore, 
the analysis of non-Bayesian errors also presents
significant interests in studies.  

The Fano's lower bound will be effective for all classifications.
The bound is general and independent of 
error type and information about $p_1$ and $p_2$. 
If no information is given about $p_1$ and $p_2$,
we obtain a ``{\it 
general upper bound}'' for non-Bayesian errors in the form:
\begin{equation}
\label{equ:8}
P_E \leq 1-H^{-1}(e)=1-G_1(H(T|Y)),
\end{equation}
which is a mirror of Fano's lower bound with mirror axis along $P_E=0.5$.
If one has information about $p_1$ and $p_2$, the
analytical upper bound of $P_E$ is
\begin{equation}
\label{equ:9}
\begin{array}{r@{\quad}l}
& \qquad \qquad P_E \leq  G_2(H(T|Y)), \\
& for ~ H(T|Y) \leq H(T|Y)_{max} ~ and ~ P_E \in [0,0.5],
\end{array} 
\end{equation}
where $H(T|Y)_{max}$ is the ``{\it upper bound of $H(T|Y)$}''
and calculated from:
\begin{equation}
\label{equ:10}
\begin{array}{r@{\quad}l}
&  H(T|Y)_{max} = H(e=p_{min}). \\
\end{array} 
\end{equation}
The analytical upper bound described in 
(14) also forms a ``{\it mirrored analytical upper bound}'', which 
will be effective for $P_E \in [0.5, 1.0]$.

From the graph of ``$P_E$ vs. $H(T|Y)$'' (Fig. 2), 
observations for non-Bayesian errors 
can also be summarized as follows:

I. In general, if no information exists about $p_1$ and $p_2$, the admissible area is 
formed by Fano's lower bound, its mirrored bound, and the 
axis of $P_E$, that is, the two-curve-one-line 
boundary ``$O-A-D-O$'' in Fig. 2. This area 
covers all other admissible areas formed from analytical bounds
for which information about $p_1$ and $p_2$ is applied. 

II. If $p_1$ and $p_2$ are known, 
the admissible area will be formed from the analytical upper
bound, its mirrored bound, and the upper bound $H(T|Y)_{max}$.
The area is controlled by $p_{min}$. For example,
if $p_{min}=0.2$, the area is enclosed by the 
four-curve-one-line boundary ``$O-F'-F-D-A'-O$'' in Fig. 2.
However, if $p_1 = p_2=0.5$,
two admissible areas are specifically formed. Their two-curve boundaries
are ``$O-F'-A-O$'' and ``$D-F-A-D$'', respectively.

III. All admissible areas, whether with or without 
information of $p_1$ and $p_2$,
are closed. The areas are formed differently with respect to
the given information. The more information available,
the tighter the bounds become, or the smaller the 
admissible areas become. In general, non-Bayesian error $P_E$ can be 
higher than Kovalevskij's bound.

{\it General upper bound of non-Bayesian errors}: 
For non-Bayesian classifications, eq. (5) with condition
$P_E=e_1+e_2>0.5$
describes a general classification setting to represent 
the general upper bound. 
Two specific settings can be obtained for demonstrations. 
One setting is described
by eq. (17) with $p_1<p_2$. The other setting is
\begin{equation}
\label{equ:11}
\begin{array}{r@{\quad}l}
& p_{11}=0.5- P_E / 2 , ~  p_{12}=P_E / 2 ,  \\
& p_{21}=P_E / 2 , ~~~~~~~~  p_{22}=0.5-P_E / 2 . 
\end{array} 
\end{equation}

{\it Mirrored analytical upper bound}: 
A mirrored analytical upper bound
is formed for non-Bayesian error
with the condition that $p_1$ and $p_2$ are known.
This bound in fact
serves as a lower bound for $P_E \in [0.5, 1.0]$.
Suppose $p_1>p_2$, 
a specific setting in classifications can be found for 
representing the mirrored 
bound: 
\begin{equation}
\label{equ:12}
\begin{array}{r@{\quad}l}
& p_{11}=p_1-e_1 , ~  p_{12}=e_1 = e > 0.5,  \\
& p_{21}=p_2 , ~~~~~~~  p_{22}=0.
\end{array} 
\end{equation}

\begin{figure}
\centering
\includegraphics[width=3.3in]{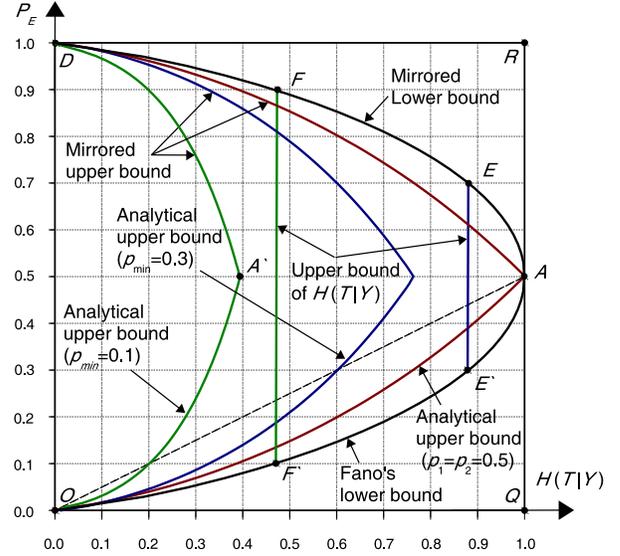}
\caption{Plot of ``$P_E$ vs. $H(T|Y)$'' giving 
the analytical bounds and the mirrored bounds.  
}
\label{fig:2}
\end{figure}

\section{Interpretations to some key points}

Further interpretations are given to
the key points shown in Fig. 1 and Fig. 2.
Those key points may hold special features in classifications. 

{\it Point O}: 
This point represents a zero value of $H(T|Y)$. 
It also suggests a ``{\it perfect classification}'' without
any error ($P_e=P_E=0$) by a specific setting 
of the joint probability:
$$\setlength\arraycolsep{0.1em}
 \begin{array}{r@{\quad}l}
& p_{11}=p_1, ~  p_{12}=0,  \\
& p_{21}=0, ~~  p_{22}=p_2.
\end{array} 
\eqno (24) 
$$

{\it Point A}: 
This point represents maximum ranges of $H(T|Y)=1$ for ``{\it class-balanced}''
classifications ($p_1=p_2$). Three specific classification settings 
can be obtained for representing this point. The two settings are
actually ``{\it no classification}'': 
$$\setlength\arraycolsep{0.1em}
 \begin{array}{r@{\quad}l}
& p_{11}=1/2, ~  p_{12}=0,  ~or ~~  p_{11}=0, ~  p_{12}=1/2,\\
& p_{21}=1/2, ~  p_{22}=0, ~~~~~~ p_{21}=0, ~  p_{22}=1/2.
\end{array} 
\eqno (25) 
$$
The other one is a ``{\it random guessing}'':
$$\setlength\arraycolsep{0.1em}
 \begin{array}{r@{\quad}l}
& p_{11}=1/4, ~  p_{12}=1/4,  \\
& p_{21}=1/4, ~  p_{22}=1/4. 
\end{array} 
\eqno (26) 
$$

{\it Point D}: 
This point occurs for non-Bayesian classifications 
in a form of:
$$\setlength\arraycolsep{0.1em}
 \begin{array}{r@{\quad}l}
& p_{11}=0, ~~  p_{12}=p_1,  \\
& p_{21}=p_2, ~  p_{22}=0.
\end{array} 
\eqno (27) 
$$
In this case, one can exchange the labels for a perfect classification.

{\it Points B (or C) and $B'$ (or $C'$)}: 
Suppose $p_1 > p_2$. The specific setting is:
$$\setlength\arraycolsep{0.1em}
 \begin{array}{r@{\quad}l}
& p_{11}=p_1-p_2, ~  p_{12}=p_2,  \\
& p_{21}=0, ~~~~~~~~  p_{22}=p_2,
\end{array} 
\eqno (28) 
$$ 
for Point $B$ when $p_2=0.4$ (or Point $C$ when $p_2=0.2$), 
and two specific settings for Point $B'$ (or Point $C'$) are:
$$\setlength\arraycolsep{0.1em}
 \begin{array}{r@{\quad}l}
& p_{11}=p_1, ~  p_{12}=0,\\
& p_{21}=p_2, ~  p_{22}=0, 
\end{array} 
\eqno (29) 
$$
or 
$$\setlength\arraycolsep{0.1em}
 \begin{array}{r@{\quad}l}
& p_{11}=0.5-p_2/2, ~  p_{12}=p_2/2,\\
& p_{21}=p_2/2, ~~~~~~~~  p_{22}=0.5-p_2/2.
\end{array} 
\eqno (30) 
$$

{\it Points E (or F) and $E'$ (or $F'$)}: 
Suppose $p_1 > p_2$. The specific setting is: 
$$\setlength\arraycolsep{0.1em}
 \begin{array}{r@{\quad}l}
& p_{11}=0, ~ p_{12}=p_1,  \\
& p_{21}=0, ~ p_{22}=p_2,
\end{array} 
\eqno (31) 
$$
for Point $E$ when $p_2=0.3$ (or Point $F$ when $p_2=0.1$), 
and eq. (30) for Point $E'$ (or $F'$)
on the given value of $p_2$.

{\it Point $A'$}: 
Suppose $p_1 > 0.5$. The specific setting for Point $A'$ is: 
$$\setlength\arraycolsep{0.1em}
 \begin{array}{r@{\quad}l}
& p_{11}=p_1-0.5, ~ p_{12}=0.5,  \\
& p_{21}=0, ~~~~~~~~~ p_{22}=p_2.
\end{array} 
\eqno (32) 
$$

{\it Points Q and R}: 
The two points are specific due to their positions in the diagrams. 
For both types of errors, they are all considered to be
``{\it non-admissible points}'' in the diagrams, because no setting exists
in binary classifications which can represent the points.

\section{Final remarks}
\label{sec:6}
This work investigates into analytical bounds between
entropy and error probability. Two specific
schemes are applied in the theoretical derivation. 
One scheme is the utilization of joint probability distributions,
on which more general interpretations 
can be obtained for understanding the bounds.
The other scheme is the 
closed-form solution of the maximization or
minimization to the related functions.  
We derived the analytical bounds for both
types of Bayesian errors and non-Bayesian ones.
While a new interpretation is given to 
Fano's lower bound, the analytical upper bounds
are achieved which show tighter than Kovalevskij's upper bound.

To emphasize the importance of the study, we present discussions
below on the selection of learning targets 
between error and entropy from the perspective of machine learning. 
The analytical bounds derived in this work 
provide a novel solution to link both
learning targets in the related studies.
Error-based learning is more conventional
because of its compatibility with our intuitions
in daily life, such as ``{\it trial and error}''. 
Significant studies have been reported
under this category. In comparison,
information-based learning \cite{principe2010}
is new and uncommon in applications, such as classifications.
Entropy is not a well-accepted concept related to our
intuition in decision making. This is one of the reasons
why the learning target is chosen mainly based on error,
rather than on entropy. However, we consider that error is
an empirical concept, whereas entropy is generally more theoretical.
In \cite{hu2012a}, we demonstrated that 
entropy can deal with both concepts of ``{\it error}'' and ``{\it reject}''
in abstaining classifications. 
Information-based learning \cite{principe2010} presents 
a promising and wider perspective for exploring and interpreting
learning mechanisms. 

When considering all sides of the issues stemming 
from machine learning studies, we believe that 
``{\it what to learn}'' is a primary problem. However, 
it seems that more investigation is focused on the issue
of ``{\it how to learn}''.
Moreover, in comparison with the long-standing yet hot theme of ``{\it feature selection}'', 
little study has been done from the perspective of ``{\it learning target selection}''.
We propose that this theme should be emphasized in the study of
machine learning. Hence, the relations studied in this work 
are very important and crucial to the extent that researchers, 
using either error-based or entropy-based approaches, are able 
to reach a better understanding about its counterpart.  

\appendices
\section{Proofs of the analytical bounds}

For a binary classification, a closed-form relation
of conditional entropy and error probabilities is derived
from the joint probability (5):
$$\setlength\arraycolsep{0.1em}
 \begin{array}{r@{\quad}l}
 H(T|Y)  = & H(T)-MI(T,Y)\\
 = & -p_{1}log_2 p_1-p_{2}log_2 p_2 \\
& -{e_1}log_2 \frac{e_1}{(p_2+e_1-e_2)p_1} \\
& -{e_2}log_2  \frac{e_2}{(p_1-e_1+e_2)p_2} \\ 
& -{(p_1-e_1)}log_2  \frac{(p_1-e_1)}{(p_1-e_1+e_2)p_1} \\ 
& -{(p_2-e_2)}log_2  \frac{(p_2-e_2)}{(p_2+e_1-e_2)p_2}.
\end{array} 
\eqno (A1) 
$$

Based on eq. (A1), the analytical functions of
lower bound and upper bound should be derived from the following 
definitions, respectively:
$$\setlength\arraycolsep{0.1em}
 \begin{array}{r@{\quad}l}
& G_1^{-1}(e, p_{min})  = arg \max \limits_{e}H(T|Y).
\end{array} 
\eqno (A2) 
$$
$$\setlength\arraycolsep{0.1em}
 \begin{array}{r@{\quad}l}
& G_2^{-1}(e, p_{min})  = arg \min \limits_{e}H(T|Y) .
\end{array} 
\eqno (A3) 
$$
The meanings of lower and upper are exchanged in (A2) and (A3) 
respectively, because the 
input variable is $e$ in the derivations. 
A single independent parameter is given to 
$p_{min}$, which is assumed to be known in the derivations. 

However, in the background of binary classifications, 
the function $H(T|Y)$ in (A1) has  
two independent variables, $e_1$ and $e_2$.
This feature causes a difficulty in the direct 
derivation of (A2) or (A3) based on a single variable function $e$.
The difficulty is the multiple solutions of $e_1$ and $e_2$ to the same bound,
which makes the derivation to be tedious. 
For overcoming this difficulty,  
we adopt Maple\texttrademark 9.5
(a registered trademark of Waterloo Maple, Inc.) for implementing the
derivations. Using the Maple
code shown in Appendix B, one is able to confirm the
derivations easily for the multiple-to-one relations of the bound. 

\begin{proof} {\it On the analytical lower bound function $G_1^{-1}(e, p_{min})$}:

From information theory \cite{cover2006}, one can have the following
conditions for mutual information:
$$\setlength\arraycolsep{0.1em}
 \begin{array}{r@{\quad}l}
& 0 \leq  MI(T,Y) \leq H(T)=H(e). \\
\end{array} 
\eqno (A4) 
$$
Hence, eq. (A1) describes that, when $MI(T,Y)=0$, one can have the
maximum results of $H(T|Y)$. We can show that
the generic classification setting in eq. (16)
will result in the condition of $MI(T,Y)=0$. Using the Maple
code,
one can substitute either condition from (16) into (A1), 
and always arrive at the same results on $MI(T,Y)=0$
and the analytical lower bound function in terms of $e$ and $p_{min}$. 
\end{proof}

\begin{proof} {\it On the analytical upper bound function $G_2^{-1}(e, p_{min})$}:

Eq. (A1) suggests that the maximum solution of $MI(T,Y)$ should be
equivalent. For achieving a single-variable function in (A3),
we need to solve the following problem first: 
$$\setlength\arraycolsep{0.1em}
 \begin{array}{r@{\quad}l}
& e = arg \max \limits_{given ~ e_2}MI(T,Y),
\end{array} 
\eqno (A5) 
$$
where $MI$ is described implicitly by two independent variables $e$ and $e_2$.
Due to high complexity of the nonlinearity in $MI$, we are unable
to obtain the direct relation between $e$ and $e_2$.
Therefore, we solve the problem of (A5) by examining the 
differential function of $MI(T,Y)$ 
with respect to $e$:
$$\setlength\arraycolsep{0.1em}
\begin{array}{r@{\quad}l}
& \frac {d} {de} MI(T,Y) =  log_2 ( \frac {(1-p_2-e+ 2e_2)} {(1-p_2-e+e_2)} 
\frac {(e-e_2)} {(e-2e_2+p_2)} ) ,
\end{array} 
\eqno (A6) 
$$
where we consider $e_2$ and $p_2$ as the constants. 
Suppose the condition $1 > p_1 > p_2 > e \geq e_2 \geq 0$,
one can prove that (A6) is always negative and without singularity.
Hence, $MI(T,Y)$ is a monotonously decreasing function 
with respect to $e$ for the given condition.
The maximum $MI(T,Y)$ will require the smallest $e$.
From $e=e_1+e_2$ and the given $e_2$
in (A5), one can derive the solutions $e=e_1$ and 
$e_2=0$. The specific classification setting associated to 
the solutions is shown in (18).
For the other conditions with the same value of $e$,   
one can always obtain the same value on the maximum of $MI(T,Y)$. 
The analytical upper bound function will be always the same 
in terms of $e$ and $p_{min}$.
\end{proof}

The proof of mirrored bounds can be obtained directly in the similar
principle, and is neglected here.   

\begin{figure*}[!t]
\section{Maple code for the derivations}
\begin{verbatim}
> #  Maple code for deriving the analytical lower bound
> restart; # Clean the memory     
> # Shannon entropy                               
> HT:=-p1*log[2](p1)-p2*log[2](p2);  
> # Terms of joint probability             
> p11:=(p1-e1);p12:=e1;p22:=p2-e2;p21:=e2;    
> # For the generic setting in (16) 
> e1:=p1*(p2-e2)/p2;p1:=1-p2;       
> # Intermediate variables             
> q1:=p11+p21;q2:=p12+p22;   
> # Mutual information                                              
> MI:=p11*log[2](p11/q1/p1)+p12*log[2](p12/q2/p1);
> MI:=MI+p22*log[2](p22/q2/(1-p1))+p21*log[2](p21/q1/(1-p1));   
> MI:=simplify(MI,ln);    # Solution of mutual information
                            MI := 0    
> # The analytical lower bound function
> HTY:=simplify((HT-MI),ln); 
> # Display of the lower bound function in terms of e and p2
                     (1 - p2) ln(1 - p2)   p2 ln(p2)
            HTY := - ------------------- - ---------
                            ln(2)            ln(2)              
> #  Maple code for deriving the analytical upper bound
> restart; # Clean the memory       
> # Shannon entropy                               
> HT:=-p1*log[2](p1)-p2*log[2](p2);  
> # Terms of joint probability             
> p11:=(p1-e1);p12:=e1;p22:=p2-e2;p21:=e2;  
> # For error variable  
> e1:=e-e2;p1:=1-p2;
> # Intermediate variables             
> q1:=p11+p21;q2:=p12+p22;   
> # Mutual information                                             
> MI:=p11*log[2](p11/q1/p1)+p12*log[2](p12/q2/p1);
> MI:=MI+p22*log[2](p22/q2/(1-p1))+p21*log[2](p21/q1/(1-p1));  
> MI_dif:=simplify(combine(diff(MI,e),ln, symbolic)); 
> # Display of diffential function of MI in (A6)
                    /  (-1 + p2 + e - 2 e2) (e - e2)   \
                  ln|----------------------------------|
                    \(-1 + p2 + e - e2) (e - 2 e2 + p2)/
        MI_dif := --------------------------------------
                                  ln(2)                 
> # For the generic setting in (18) 
> e1:=e;e2:=0;p1:=1-p2;       
> # Intermediate variables             
> q1:=p11+p21;q2:=p12+p22;                         
> # Mutual information                                             
> MI:=p11*log[2](p11/q1/p1)+p12*log[2](p12/q2/p1);
> # Neglect one term below from the entropy definition of 0*log(0)=0
> MI:=MI+p22*log[2](p22/q2/(1-p1));   
> # The analytical upper bound function
> HTY:=combine(simplify(combine(simplify(HT-MI),ln,symbolic)));  
> # Display of the upper bound function in terms of e and p2   
                       /e + p2\       /e + p2\
                  p2 ln|------| + e ln|------|
                       \  p2  /       \  e   /
            HTY:= ----------------------------
                             ln(2)                            
\end{verbatim}

\end{figure*}

\section*{Acknowledgments}
This work is supported in part by NSFC (No. 61075051
for BG and No. 60903089 for HJ).

%

\bibliographystyle{IEEETrans}


%





\end{document}